\documentclass[twocolumn,floatfix, nofootinbib,prd,preprintnumbers,showpacs,showkeys,superscriptaddress,longbibliography]{revtex4-2}
\usepackage{slashed}
\usepackage{amsmath}
\usepackage{booktabs}
\usepackage{array}

\usepackage{graphicx} 
\usepackage{mathrsfs}
\usepackage{amssymb}
\usepackage{amsmath}
\date{\today}
\usepackage{mathtools}
\usepackage{comment}

\usepackage[mathscr]{euscript}
\usepackage{amsmath}
\usepackage{graphicx}
\usepackage{dcolumn}
\usepackage{bm}
\usepackage{epsfig}
\usepackage{amssymb,latexsym,mathrsfs}
\usepackage{graphicx}
\usepackage{color}
\usepackage{hyperref} 
\usepackage{float}
\usepackage{diagbox}

\usepackage{textgreek}

\usepackage{tikz}

\usepackage{amsmath}
\usepackage{physics}
\usepackage{csquotes} 

\hypersetup{
    colorlinks=true,
    linkcolor=red,
    citecolor=blue,
} 

\usepackage{amsmath}
\usepackage{amssymb}
\usepackage{subfigure}
\usepackage{url}
\usepackage{xcolor}
\usepackage{color}
\usepackage{soul}
\definecolor{amaranth}{rgb}{0.9, 0.17, 0.31}
\definecolor{purple(munsell)}{rgb}{0.62, 0.0, 0.77}
\definecolor{americanrose}{rgb}{1.0, 0.01, 0.24}
\definecolor{palatinateblue}{rgb}{0.15, 0.23, 0.89}
\definecolor{royalblue(web)}{rgb}{0.25, 0.41, 0.88}
\definecolor{hanpurple}{rgb}{0.32, 0.09, 0.98}
\definecolor{beaublue}{rgb}{0.74, 0.83, 0.9}
\definecolor{carminered}{rgb}{1.0, 0.0, 0.22}
\definecolor{brightpink}{rgb}{1.0, 0.0, 0.5}
\definecolor{vividviolet}{rgb}{0.62, 0.0, 1.0}
\hypersetup{ linktoc=all,
    colorlinks, linkcolor={palatinateblue},
    citecolor={brightpink}, urlcolor={amaranth}}

\usepackage{comment}

\newcommand{\be}{\begin{equation}}
\newcommand{\ee}{\end{equation}}
\newcommand{\bs}{\begin{split}} 
\newcommand{\bea}{\begin{eqnarray}}
\newcommand{\eea}{\end{eqnarray}}

\newcommand{\lcdm}{$\Lambda$CDM}

\newcommand{\omw}{\Omega_w}

\newcommand{\bes}{\begin{subequations}}
\newcommand{\ees}{\end{subequations}}

\newcommand{\bo}{\raise-1mm\hbox{\Large$\Box$}}

\usepackage{cleveref}
\crefname{equation}{Equation}{Equations}
\Crefname{equation}{Equation}{Equations}

\begin{document}

\title{Cosmic Acceleration from Nothing}
\author{Michael R.R. Good}
\email{michael.good@nu.edu.kz}
\affiliation{Physics Department \& Energetic Cosmos Laboratory, Nazarbayev University,\\
Astana 010000, Qazaqstan}
\affiliation{Leung Center for Cosmology and Particle Astrophysics,
National Taiwan University,\\ Taipei 10617, Taiwan}
\author{Eric V.\ Linder}
\email{evlinder@lbl.gov} 
\affiliation{Berkeley Center for Cosmological Physics \& Berkeley Lab, University of California,\\ Berkeley CA 94720, USA.}

\begin{abstract} 
We demonstrate that if the universe started as a vacuum 
fluctuation rather than from a singular Big Bang state, the 
universe must have a late-time cosmic acceleration. 
This is required by a ``cosmological sum rule'' derived using 
the Schwarzian form of the Friedmann equations. We discuss 
possible connections to conformal and M\"obius transformations, and also 
compute that the best fit present cosmic data is consistent with the necessary crossing 
of the Schwarzian through zero having occurred 
(while it would not yet have happened in 
a \lcdm\ cosmology). 
\end{abstract} 

\date{\today} 

\maketitle

\section{Introduction} 

Cosmic acceleration may arise from a vacuum energy, i.e., \  
a cosmological constant or a new scalar field component. 
We do not know any reason why these would {\it have\/} to 
exist and cause the observed late-time acceleration of the 
cosmic expansion. 

Here, we explore changing the vacuum structure in the early 
universe so that the universe arises as a quantum 
fluctuation in empty space rather than directly from a 
hot, dense Big Bang and how this might enforce a late-time acceleration. 

In Section~\ref{sec:expan}, we introduce the Schwarzian derivative in cosmology and explore its implications for the universe's evolution. Section~\ref{sec:schw} discusses the Schwarzian's connection to fundamental symmetries, such as Lorentz and M\"obius transformations, its use in conformal field theory and gravitational dynamics, and motivation for exploring its role in cosmology. 
Section~\ref{sec:sum} uses the Schwarzian to derive a cosmological sum rule, which imposes a necessary condition on the universe’s expansion history. Leveraging this sum rule, we compute constraints on cosmic acceleration in Section~\ref{sec:cosmicacc}, showing that late-time acceleration follows from a vacuum fluctuation origin, and compare to current data. 
Finally, Section~\ref{sec:concl} summarizes the main findings.

\section{Cosmic Expansion} \label{sec:expan} 

The Schwarzian derivative plays a key role in many fields 
of physics, including connections to Lorentz transformations 
and conformal field theory (see \Cref{sec:schw}). Here we 
will apply it to standard 
cosmology, revealing intriguing implications connecting the 
early and late universe. 

Reviewing the standard quantities of cosmic expansion, we start with 
the Robertson-Walker metric of a homogeneous, isotropic 
spacetime, 
\be 
ds^2=-dt^2+a^2(t)\,\left[\frac{dr^2}{1-kr^2}+r^2(d\theta^2+\sin^2\theta d\phi^2)\,\right]\ , 
\ee 
where $a(t)$ is the expansion factor, $k$ the spatial 
curvature constant, and $\{t,r, \theta,\phi\}$ are the 
coordinates. Using the Einstein field equations of 
general relativity yields the Friedmann equations determining 
the expansion behavior $a(t)$ in terms of the energy-momentum 
contents (and $k$). 

Those equations of motion can be written in terms of 
combinations of derivatives of the expansion factor, e.g.,\ 
\bea 
H(a)&\equiv& \frac{\dot a}{a}\ , \\  
q(a)&\equiv& -\frac{a\ddot a}{\dot a^2}\ , 
\eea 
called the Hubble parameter $H$, or logarithmic expansion 
rate, and the deceleration parameter $q$. While third 
derivatives do not enter the field equations directly, 
it will be useful to define the jerk parameter, 
\be 
j(a)\equiv \frac{a^2\dddot a}{\dot a^3}\ . 
\ee 

Now we introduce the Schwarzian, 
denoted here by brackets, 
\be 
\{a,t\}\equiv \frac{\dddot{a}}{\dot{a}} - \frac{3}{2} \left(\frac{\ddot{a}}{\dot{a}}\right)^2\ . \label{eq:szdef} 
\ee
This can be written in terms of the previously defined 
expansion quantities as (e.g.\ 
\cite{Gibbons:2014zla}), 
\be 
\{a,t\} = H^2\,\left(j-\frac{3}{2}\,q^2\right)\ . 
\ee 
We note that this has several important properties, 
both cosmological -- it only involves $a(t)$ and has no 
explicit dependence on spatial curvature $k$ -- 
and involving physical symmetries (discussed in the next 
section). 

Before proceeding, let us connect these quantities to 
simple Friedmann cosmologies. For energy-momentum contributions 
from noninteracting components, each with dimensionless 
energy density (as a fraction of the critical energy density) 
$\Omega_w$ and equation of state (pressure to energy density)  
ratio $w$, we have 
\be 
q=\frac{1}{2}\sum (1+3w)\omw\ . 
\ee 
We can see explicitly that spatial curvature 
does not enter into $q(a)$, since it 
effectively has $w_k=-1/3$. That is, 
$\Omega_k\equiv -ka^{-2}/(H_0^2)$, where $H_0=H(a=1)$ is the 
expansion rate today, follows the standard  
energy density dilution with expansion as $a^{-3(1+w)}=a^{-2}$. 

Furthermore, the jerk, 
\be 
j=q+2q^2-q'\ , 
\ee 
where prime denotes $d/d\ln a$, 
is also immune to spatial curvature. For 
illustration we write here the expressions for the case of matter 
(which has $w=0$) plus one additional component, for a critical density universe (so 
$\Omega_m+\omw=1$). Then 
\begin{align}
q&= \frac{1}{2}+\frac{3}{2}w\omw\ , \\ 
j&= 1+\frac{9}{2}w(1+w)\omw-\frac{3}{2}w'\omw\ , \\ 
\{a,t\}&= \frac{H^2}{8} \left[5+18w(1+2w)\omw-12w'\omw-27w^2\omw^2\right]. \label{eq:schinw} 
\end{align}  
Note that when the additional component is a cosmological 
constant ($w=-1$), called \lcdm\ cosmology, 
then $q=(1-3\omw)/2$, ranging from $1/2$ 
in matter domination ($\omw=0$) to $-1$ in cosmological 
constant domination ($\omw=1$, or de Sitter asymptote). 
However, $j=1$ for all times for \lcdm.

\section{Schwarzians and Symmetries} \label{sec:schw} 

The Schwarzian derivative as defined in \Cref{eq:szdef} has 
close connections with the symmetry group $SL(2,R)$ and hence 
Lorentz transformations, conformal transformations, and 
M\"obius transformations (see e.g.\ \cite{Kitaev:2017hnr}). 
The symmetry gives invariance under physical actions like 
boosts, dilations, and rotations. 

Among other uses in physics, the Schwarzian appears in conformal field theory (CFT) and AdS$_2$ gravity, governing boundary dynamics in Jackiw-Teitelboim (JT) gravity (see e.g.\ \cite{Mertens:2022irh,Mertens:2018fds}). It controls the low-energy sector of the SYK model, a maximally chaotic system (see e.g. \cite{Maldacena:2016hyu,Qi:2018rqm}), by describing the soft reparametrization modes that emerge from the breaking of conformal symmetry. 
It measures deviations breaking M\"obius symmetry and 
plays a role in 
conformal transformations in cosmology \cite{Gibbons:2014zla}. 
In black hole evaporation analogs, such as the moving mirror model \cite{DeWitt:1975ys, Davies:1976hi}, it determines the energy flux (the stress tensor) of quantum particles radiated by an accelerating mirror (see e.g.\ \cite{Birrell:1982ix, Fabbri:2005mw}). Its relation to M\"obius transformations 
plays a central role in analyzing invariant energy flux, particle spectrum, and entanglement entropy of de Sitter and black hole horizons \cite{Good:2021iny}. 

Given the usefulness of the Schwarzian, its appearance 
relating basic quantities of cosmic expansion in 
\Cref{eq:szdef} merits 
investigation of possible implications.

\section{Cosmological Sum Rule} \label{sec:sum} 

We find that employment of the Schwarzian 
reveals a constraint known in physics as a sum rule, 
or integral constraint, here applied to the cosmic 
expansion. Let us parallel the use of the Schwarzian 
in the moving mirror case (also known as an accelerating 
boundary correspondence to a spacetime or black hole 
horizon). There, it proved that a unitary solution (and hence 
an evaporating black hole without information loss) 
gave a horizon sum rule (in that case, requiring some 
epoch where energy flux was negative) 
\cite{Good:2019tnf}. In our cosmological  
case, we will find a sum rule that implies that the universe 
must have some epoch of accelerating expansion. 

In \cite{Good:2019tnf,Good:2021iny} the Schwarzian and 
the resulting sum rule appeared most clearly with use 
of the rapidity, as in special relativity $\eta=\tanh^{-1}(dx/dt)$, but in spacetime null coordinates 
$\{u,v\}=\{t-x,t+x\}$ so 
\be 
\eta=\frac{1}{2}\,\ln\frac{dv}{du}\ .  
\ee 
Therefore we define 
\be 
\chi \equiv \frac{1}{2}\,\ln \dot{a}\ , 
\ee 
and stay restricted to an expanding universe, $\dot a>0$. 

We can now write the Schwarzian as 
\be 
\{a,t\} = \frac{1}{2}\,\left[\ddot{\chi} - \dot{\chi}^2\right] = \frac{1}{2}\,e^{\chi}\,\frac{d}{dt}\left(e^{-\chi}\dot{\chi}\right)\ . 
\ee 
Removing the (always positive) prefactor and integrating 
over time gives a total derivative, 
\bea  
\int_{0}^{+\infty} dt\; \dot{a}^{-1/2}\; \{a,t\} = \left.\left(\dot{a}^{-1/2} \frac{\ddot{a}}{\dot{a}}\right)\right|_{t=0}^{t=+\infty}\ . \label{eq:integ} 
\eea 
Suppose that in the future $a\sim t^n$ as $t\to+\infty$. Then the 
quantity in parentheses goes as $n^{-1/2}(n-1)t^{-(1+n)/2}$, 
and so for expansion ($n>0$) the end point term will vanish. 
(It will also vanish for a de Sitter state, $a\sim e^{Ht}$.) 
At early times, the boundary term will vanish if $\ddot a=0$, 
i.e.\ constant $\dot a$, 
hence constant $\chi$, or $a\sim t$. 

Thus, we have a cosmological sum rule, 
\be 
\int_{0}^{+\infty} dt\; \dot{a}^{-1/2} \{a,t\} = 0\ , 
\ee 
if the early universe began as a Milne universe, a state 
empty of all contents (and with negative curvature $k$), having $a\sim t$. 
Note that the Milne universe is, with a coordinate 
transform, a patch of Minkowski spacetime (which 
has $k=0$), also a 
vacuum solution\footnote{One 
could also regard Minkowksi spacetime as 
the limit $h\to0$ in $a\sim e^{ht}$, i.e.\ 
a static limit. In this case Minkowski contributes $h^{1/2}e^{-ht/2}$ to the 
integral Eq.~(\ref{eq:integ}), which vanishes as $h\to0$, as 
required.}.   
In either case, if the universe begins not with a hot, dense 
Big Bang state but as a quantum fluctuation about empty 
spacetime, then the cosmological sum rule holds. 

Since $\dot{a}^{-1/2}=e^{-\chi}>0$ then 
the Schwarzian $\{a,t\}$ must cross zero to 
satisfy the sum rule, and hence 
be negative for some time during its evolution. 
Note also that since for a 
Milne universe the Schwarzian is zero 
(indeed $q=0=j$), then we can extend the 
lower limit of the integral indefinitely, 
if the universe arose from a Milne state.

\section{Cosmic Acceleration} 
\label{sec:cosmicacc} 

The results of the previous section are completely general, requiring only expansion and vanishing boundary terms, and do not 
rely on any assumptions about splitting 
energy density into matter plus dark energy. They imply 
that $j-(3/2)q^2$ must cross zero. This does 
not mean that $j(a)=(3/2)q^2(a)$ for all 
times (for which the only expanding solution is $a\sim t$, e.g.\ an eternal Milne 
universe), but rather that the Schwarzian must 
cross zero. 

To give a flavor of this, and be quantitative, in this section we will 
explore the case of matter plus some other 
component, as in 
\Cref{eq:schinw}, initially with constant $w$. 

First, let's consider a matter plus cosmological 
constant (\lcdm) universe. Then $(8/H^2)\{a,t\}=5+18\omw-27\omw^2$. As the universe evolves from early 
to late time, $\omw$ increases from 0 to 1, and the right-hand side varies from 5 to $-4$, indeed crossing zero. 
Next, we can allow the component adding to matter to have 
an arbitrary constant $w$. In this case, at early times ($\omw\to0$) 
the Schwarzian will always be positive. At late times ($\omw\to1$) 
we have an equation for the crossing given by $(5+3w)(1+3w)=0$. 
When $w>-1/3$ no crossing can occur, i.e.\ we cannot 
satisfy the cosmological sum rule. 

Thus, if the universe 
starts as a quantum fluctuation from an empty, Milne state, 
it must have late-time acceleration ($w<-1/3$, $\omw\to1$)! 

However, for $w<-5/3$ the Schwarzian stays positive at 
late times and never crosses zero. Thus, we are forced by 
the sum rule to have 
\be 
-5/3<w<-1/3\ . 
\ee 
We can evaluate when the crossing occurs by solving for the roots of 
\Cref{eq:schinw} in terms of $\omw$, 
\be 
\omw^{\rm cross}=\frac{1}{3w}\,\left[1+2w-\sqrt{(1+2w)^2+5/3}\right]\ , \label{cross}
\ee 
where the other root would have $\omw<0$ for $w<0$.  

\begin{figure}[h]
    \centering
    \includegraphics[width=0.45\textwidth]{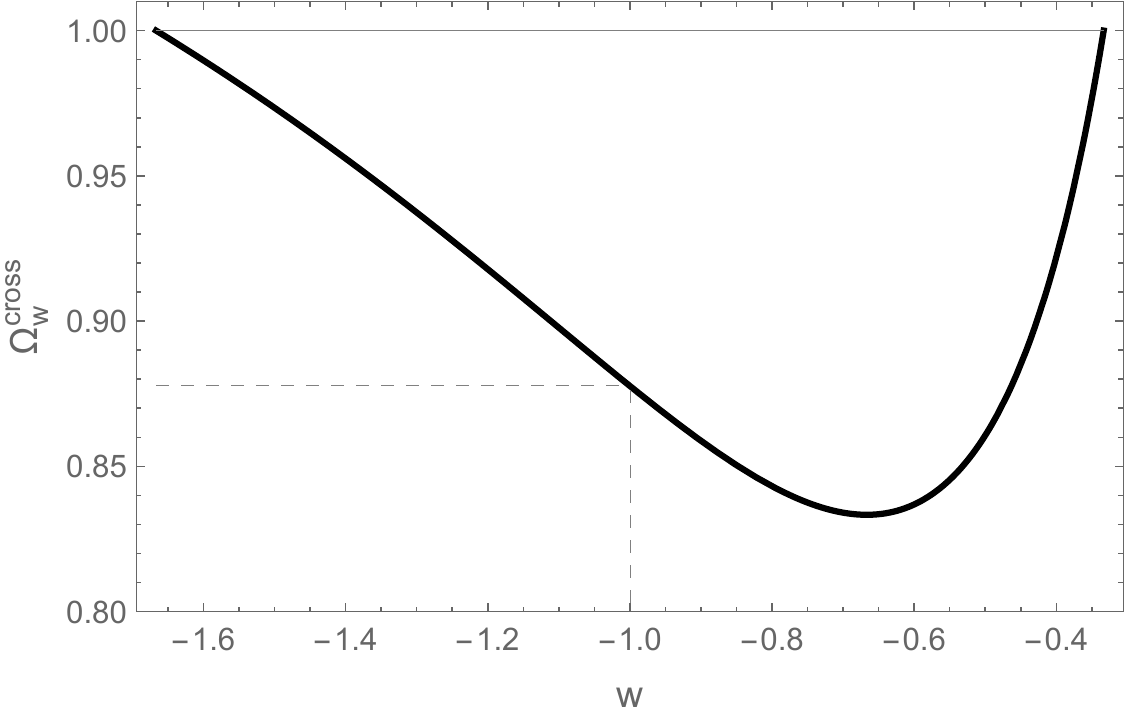}
    \caption{The sum rule requires the Schwarzian to cross zero, hence the presence of a component with $-5/3 < w < -1/3$. We plot the value of $\omw^{\rm cross}$, i.e.\ when in the expansion history this occurs, for each $w$, from \Cref{cross}. For \lcdm\ ($w=-1$), for example, the crossing will occur when $\omw=0.878$.  
    }
\label{fig:omwroot}
\end{figure}

\Cref{fig:omwroot} shows at what $\omw$ the crossing 
occurs as a function of $w$. For $w=-1$, it occurs at 
$\omw=0.878$, i.e.\ in the future in our universe. 
Outside the range $-5/3<w<-1/3$, a crossing would require 
$\omw>1$ (i.e.\ a negative matter density or positive curvature).

\begin{figure}[h]
    \centering
    \includegraphics[width=0.45\textwidth]{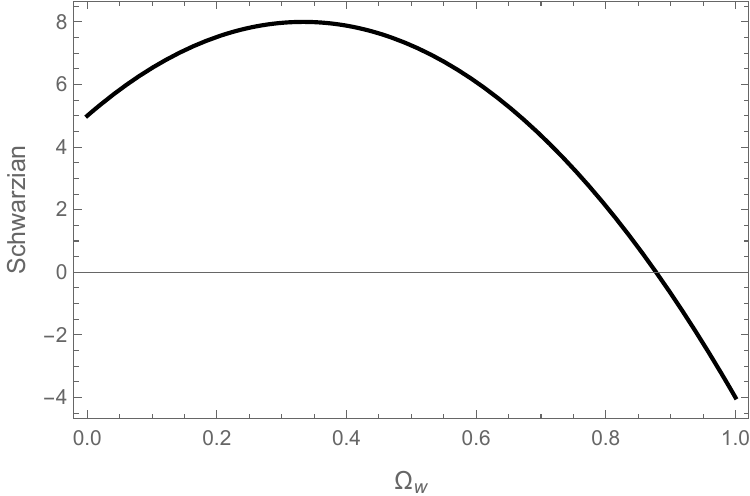}
    \caption{The Schwarzian, actually $8\{a, t\}/H^2$, is plotted 
    for \lcdm\ cosmology, showing the required zero crossing. 
    }
\label{fig:schwarz}
\end{figure}

\Cref{fig:schwarz} plots the Schwarzian (actually $8\{a,t\}/H^2$) 
for \lcdm\ cosmology. Recall that we can consider $\omw$ as a 
measure of time, going from $\omw=0$ in the early universe to 
$\omw=1$ in the late universe.

Relaxing the constancy of the equation of state $w$, we see that including the $w'$ 
term from \Cref{eq:schinw} imposes a constraint 
on the dark energy dynamics such that we require 
\be 
w'>\frac{5+18w(1+2w)\omw-27w^2\omw^2}{12\omw}\ , \label{eq:wp} 
\ee 
at some time. 
While we cannot test this for the future, 
i.e.\ all the way to $\omw=1$, we might evaluate the 
expression to 
see if our universe is at least close to satisfying 
the crossing required by the cosmological 
sum rule. As one example, in \lcdm\ we see from 
\Cref{fig:schwarz} that even 
at $\omw=0.7$ we cannot tell the crossing is 
close. However, for observational data 
results similar to DESI 
\cite{DESI:2024mwx} with a large 
$w'\sim 1$, $w\sim -0.7$ the inequality is 
satisfied today, seeming to 
indicate the crossing has already been 
accomplished, as the sum rule requires.

\section{Conclusions} \label{sec:concl} 

If the universe started from nothing -- specifically a 
Milne state -- with a quantum fluctuation generating 
matter (including radiation), then we find that at some later time the cosmic expansion must accelerate. This 
is in sharp contrast to the usual hot, dense Big Bang 
picture where the universe ``starts'' with matter and 
radiation, but there is no guarantee that expansion will 
ever accelerate -- that has to be put in by hand. 

The structure of the Friedmann equations in Einstein's 
general relativity, under the symmetry properties of 
the Schwarzian and the resulting cosmological sum rule we derived, 
do guarantee that any quantum fluctuation from a Milne state 
generates not only matter and radiation 
(or at least a component with an equation of state $w>-1/3$) 
but also an additional component with $-5/3<w<-1/3$ that will dominate at late 
times and accelerate cosmic expansion. 
The Schwarzian alone does not say how the quantum 
production process works. 

The Milne state can persist indefinitely, until the 
quantum transition, and since during it $a\sim t$ then 
there is still the possibility of a singular origin in the past, 
$a=0$ -- a Big Bang but an empty one with no matter to 
witness creation. Once the quantum fluctuation generates 
the matter, cosmic evolution can follow the standard 
hot, dense early cosmology, satisfying all 
the usual constraints of primordial nucleosynthesis, the cosmic microwave background radiation, and growth of large scale structure, but now with late-time 
acceleration guaranteed. We also note that 
nothing prevents multiple 
epochs of acceleration, with the Schwarzian $j-(3/2)q^2$ 
crossing zero several times\footnote{ 
Note that inflation is unlikely to satisfy the sum rule on its own, since the exponentially small $\dot a^{-1/2}$ in the integral means its negative Schwarzian would be quickly overcome by the positive contribution from radiation after inflation ends, in much the same way that a late-time de Sitter state still leads to a vanishing boundary term.
}. 

This is an attractive scenario, where the origin 
of the universe out of nothing, utilizing the 
symmetry properties inherent to Einstein's equations of motion, automatically provides for a late time accelerated expansion such as we observe. The prevalence of the Schwarzian and its ties to so many fields of physics offer the hope that the quantum origin itself 
may be part of a deeper theory, whether in 
conformal field theory, quantum gravity, or 
symmetry breaking. Finally, \Cref{eq:wp} offers a predictive constraint on the dark energy equation of state behavior that can be tested with precision observations.

\begin{acknowledgments}
M.G. is supported in part by the FY2024-SGP-1-STMM Faculty Development Competitive Research Grant (FDCRGP) no.201223FD8824 and SSH20224004 at Nazarbayev University in Qazaqstan. 
\end{acknowledgments}

\bibliography{main} 

\end{document}